\documentclass[aps,twocolumn,a4paper,pra,superscriptaddress,floatfix]{revtex4}%
\usepackage{amssymb}
\usepackage{amsmath}
\usepackage{amsfonts}
\usepackage{graphicx}
\usepackage{graphics}

\def\P{\mathrm{P}}    
\def\T{\mathrm{T}}    
\def\S{\mathrm{S}}    
\def\PFA{\mathrm{PFA}}
\def\TE{\mathrm{TE}}
\def\TM{\mathrm{TM}}

\def\cF{\mathcal{F}}

\def\cL{\mathcal{L}}
\def\cK{\mathcal{K}}
\def\cR{\mathcal{R}}
\def\cM{\mathcal{M}}
\def\max{\mathrm{max}}
\def\perf{\mathrm{perf}}
\def\plas{\mathrm{plas}}
\def\Drud{\mathrm{Drud}}
\def\txi{\tilde{\xi}}

\begin{document}

\title{Thermal Casimir Effect in the Plane-Sphere Geometry}

\author{Antoine Canaguier-Durand}
\affiliation{Laboratoire Kastler Brossel,
CNRS, ENS, Universit\'e Pierre et Marie Curie case 74,
Campus Jussieu, F-75252 Paris Cedex 05, France}
\author{Paulo A. Maia Neto}
\affiliation{Instituto de F\'{\i}sica, UFRJ,
CP 68528,   Rio de Janeiro,  RJ, 21941-972, Brazil}
\author{Astrid Lambrecht}
\author{Serge Reynaud}
\affiliation{Laboratoire Kastler Brossel,
CNRS, ENS, Universit\'e Pierre et Marie Curie case 74,
Campus Jussieu, F-75252 Paris Cedex 05, France}

\date{\today}

\begin{abstract}
The thermal Casimir force between two metallic plates is known to depend on the description of material properties. 
For large separations the dissipative Drude model leads to a force a factor of 2 smaller than the lossless plasma model. 
Here we show that the plane-sphere geometry, in which current experiment are performed, 
decreases this ratio to a factor of 3/2, as revealed by exact numerical and large distance analytical calculations. 
For perfect reflectors, we find a repulsive contribution of thermal photons 
to the force and negative entropy values at intermediate distances.
\end{abstract}


\maketitle

Interest in the Casimir force, arising due to the confinement of the vacuum fluctuations 
of electromagnetic fields between two reflecting bodies, has been continuously growing during
the last decade and the motivation for measuring it precisely has
led to a number of original experiments using various modern technologies
\cite{Decca07,Chan08,VanZwol08,Munday09,Jourdan09,Masuda09,deMan09}.
The Casimir force depends on a number of factors, including
the bodies' material properties \cite{Lifshitz56,Lambrecht00}, their surface state
\cite{Klim00,MaiaNeto05,LambrechtPRL08} and shape. Current experiments are performed using a
spherical and a plane surface. The force in this plane-sphere geometry
is usually calculated within the Proximity Force Approximation (PFA)
\cite{Derjaguin68} which averages the force calculated in the
plane-plane geometry over the local inter-surface distances. 
Only recently studies have been devoted to Casimir force evaluations going beyond the
domain of validity of PFA \cite{Krause07,Neto08,Gies08,Kenneth08,Emig08,CanaguierPRL09}.

The influence of temperature on the Casimir effect has given rise to
intense discussions over the last decade \cite{Bostrom00,Brevik06},
in particular because the force exhibits an unexpectedly strong correlation
with the detailed description of optical properties of the metallic
surfaces used in the experiments. 
The dielectric function of metals is often modeled by the plasma model where the plasma frequency
$\omega_\P$, depending on the density of conduction electrons, acts
as a cut-off frequency above which the reflectivity goes to zero. In
the Drude model which also accounts for the relaxation of
conduction electrons, the dielectric function at imaginary
frequencies $\omega= i\xi$ is given by $\varepsilon(i\xi) =
1+\frac{\omega_\P^2}{\xi(\xi+\gamma)}$, with the relaxation
frequency $\gamma$ related to the reduced dc conductivity
$\sigma_0 = \frac{\omega_\P^2}{\gamma}$ (see \cite{Ingold09}). The plasma
model permittivity is obtained from the Drude model one in the limit
$\gamma\to0$, while the perfect
reflector's infinite permittivity is recovered with the further limit $\omega_\P\to\infty$.

While the dissipative Drude model seems a more appropriate
description of metallic mirrors, it turns out that recent
force measurements are in better agreement with the predictions of
the lossless plasma model \cite{Decca07}.
In addition, the Casimir force between two plates at large separations turns out to be a factor of two smaller
when calculated with the Drude model compared to the one obtained with the plasma or
the perfect reflector model. This significant difference is attributed to the
vanishing contribution of TE modes at zero frequency for dissipative
mirrors entailing that for the Casimir force, contrary to the dielectric function, there is no
continuity from the Drude to the plasma model at the limit of a
vanishing relaxation \cite{Ingold09}. In contrast to the other two models, the Drude model 
also leads to negative Casimir entropy values between two plates \cite{Bezerra02}.

In the present letter, we treat plane and spherical metallic
surfaces coupled to electromagnetic vacuum and thermal fields with material properties described
by either the perfect reflector, plasma or Drude models, and show that the above mentioned features 
are considerably altered in this situation.
First, the factor of 2 between the long distance forces in Drude and plasma models is reduced to a factor 
of 3/2 decreasing even more below this value when small spheres are considered.
Second, negative entropies are found also for the perfect reflector model, 
in which case they can only be related to the plane-sphere geometry and not to dissipation. 
Finally, PFA underestimates the Casimir force within the Drude model for short distances, 
while it overestimates it at all distances for the perfect reflector and plasma model.

We consider a large parameter space generated by the five length scales involved in the problem~:
the surface separation $L$, the sphere radius $R$, the thermal wavelength $\lambda_\T=\hbar c/k_B T$ at
temperature $T$, the plasma wavelength $\lambda_\P=2\pi c/\omega_\P$
and the wavelength associated with relaxation frequency
$\lambda_\gamma =2\pi c/\gamma$.
The general scattering formula \cite{LambrechtNJP06}
for the Casimir free energy $\cF$ between a plane and a sphere is given by:
\begin{eqnarray}
\label{depart}
&&\cF=k_B T \sum_n^{'}\log{\rm det}
\left(1- \cM (\xi_n)\right) \;,\; \xi_n=\frac{2\pi n k_B T}\hbar  \nonumber \\
&&\cM(\xi_n)\equiv \cR_\S(\xi_n) e^{-\cK(\xi_n) \cL} \cR_\P(\xi_n)
e^{-\cK(\xi_n) \cL}.
\end{eqnarray}
According to (\ref{depart}), the operator $\cM$ contains the reflection operators $\cR_\S$ and
$\cR_\P$ of the sphere and the plane respectively, the latter being
evaluated with reference points placed at the sphere center and at
its projection on the plane, as well as the translation operators
$e^{-\cK(\xi_n) \cL}$ describing one-way propagation between the
reference points on a distance $\cL=L+R$; the primed sum is a sum
over the Matsubara frequencies $\xi_n$ ($n\geq0$) with the $n=0$
term counted for a half.

The upper expression is conveniently written through a decomposition
on suitable plane-wave and multipole bases.
The resulting elements of the matrix $\cR_\P$ are proportional to the Fresnel reflection coefficients
$r_p$ with $p=$TE and TM for the two electromagnetic polarizations, 
while those of $\cR_\S$ are proportional to the Mie coefficients $a_\ell,b_\ell$ \cite{Bohren83}
for electric and magnetic multipoles at order $\ell=1,2,...$, respectively.
Due to rotational symmetry around the $z$-axis, each eigenvalue of the
angular momentum $m$ gives a separate contribution to the Casimir free
energy $\cF^{(m)}$, obtained through the same formula as (\ref{depart}),
with $\cM $ reduced to the block matrix $\cM^{(m)}$ collecting the couplings
for a fixed value of $m$.
The numerical computations presented below are done after truncating the
vector space at some maximum value $\ell_\max$ of the orbital number $\ell$.
The results are thus accurate only for $R/L$ smaller than
some value which increases with $\ell_\max$, typically $R/L<10$ for
our $\ell_\max=34$, remaining a factor of 10 below current experimental values $R/L > 10^{2}$.

\begin{figure}[t]
\centering
\includegraphics[width=8cm]{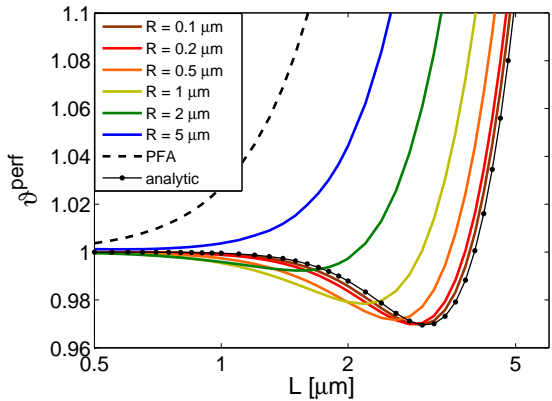}
\caption{Thermal Casimir force at $T=300$K computed between perfectly reflecting
sphere and plane, divided by the zero temperature force. Solid lines from bottom
to top correspond to increasing values of sphere radii.
The upper dashed curve represents the PFA expression while the lower dotted
curve is the analytical asymptotic expression in the $L \gg R$ limit
[Colors online].} \label{Fig1}
\end{figure}

The results of the numerical computations are shown on Fig.1 for perfect reflectors.
We have calculated the Casimir force $F^\perf$
between the plane and the sphere at ambient temperature and
then plotted the ratio $\vartheta^\perf$ of this force to its value
at zero temperature:
\begin{eqnarray}
\label{deftheta}
F^\perf(L,T)\equiv-\frac{\partial\cF^\perf}{\partial L} \quad, \quad
\vartheta^\perf\equiv\frac{F^\perf(L,T)}{F^\perf(L,0)} . &&
\end{eqnarray}
The various solid curves are drawn for different
radii $R$ of the sphere as a function of the distance $L$,
with increasing values of $R$ from bottom to top~;
the upper dashed curve on Fig.1
represents the quantity $\vartheta^\perf_\PFA$ as it would
be obtained from (\ref{deftheta}) by using PFA~;
the lower dotted curve is an analytical asymptotic expression
discussed below.

Fig.2 shows the variation of the ratio $\vartheta^\Drud$, defined as
in (\ref{deftheta}) for the Drude model with
$\lambda_\P=136$nm, $\lambda_{\gamma}/\lambda_\P=250$
and $\lambda_\T=7.6 \mu$m.
The dashed curve on Fig.2 represents $\vartheta^\Drud_\PFA$ as obtained for the Drude model by using PFA.
We do not plot the variation of $\vartheta^\plas$, defined as in
(\ref{deftheta}) for the plasma model, since it is as expected quite
close to the one shown on Fig.1 for perfect mirrors.

\begin{figure}[t]
\centering
\includegraphics[width=8cm]{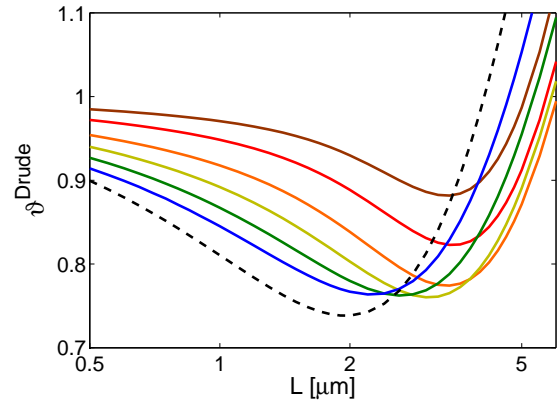}
\caption{Same plot as in Fig.1 for the Casimir force at $T=300K$
computed with the Drude model, divided by its value at zero temperature.
The dashed curve correspond to the PFA expression
[Colors online].} \label{Fig2}
\end{figure}

In most cases, the ratio $\vartheta$, starting
from unity at small distances, decreases below unity when the
distance increases, then reaches a minimum before increasing at very
large distances.
While such behavior was already observed for dissipative Drude mirrors in the plane-plane
geometry \cite{Brevik06,Bezerra02} (the dashed PFA curve on Fig.2 is also below unity for $L\lesssim 4\mu$m), 
or in the atom-surface configuration out of equilibrium \cite{Antezza05}, 
our computations show quite unexpectedly that in the sphere-plane geometry such a behavior takes place  
even for the perfect reflector and plasma models at thermodynamical equilibrium. 
Hence, for all three models the contribution of the thermal photons to the Casimir
force can be repulsive, which suggests that the entropy could be negative 
for some values of the parameters (see below).

A second important feature comes out in a striking manner from the comparison of Fig.1 and Fig.2~: 
the PFA always overestimates the effect of temperature on the force between mirrors described 
by a perfect reflector or plasma model; in contrast, it underestimates the temperature effect 
for the Drude model at small distances, while it overestimates it at large distances. 
The overestimation is however smaller than for perfect mirrors. As a consequence, for small separations 
the Drude and plasma models lead to Casimir force values much closer than predicted by PFA. 
These results clearly demonstrate the strong correlation between the effects of plane-sphere geometry,
temperature and dissipation.

In the following we will corroborate the previous numerical results by presenting analytical calculations 
of the thermal Casimir energy in the limit of large distances ($L \gg R$). 
Since the number of modes $\ell_\textrm{max}$ needed to get an accurate result decreases when $L/R$ increases, 
we may take $\ell_\textrm{max}=1$ in this limit. 
Another consequence of this limit is that the reduced frequency $\txi \equiv \xi R/c$ is very small, 
since the characteristic frequencies, which give the main contribution to the Casimir force, scale as $\xi \sim c/L$.

For perfect reflectors, where $\lambda_\P = 0$,
the dielectric function $\varepsilon$ is infinite at all frequencies and we
obtain the following low-frequency expansions for the Mie coefficients $a_1$ and
$b_1$ describing the scattering  on the sphere:
\begin{eqnarray}
\label{Mieperfect}
a_1^\perf = - \frac{2 \txi^3}{3} \quad,\quad
b_1^\perf =  \frac{\txi^3}{3}  .
\end{eqnarray}
The other steps in the calculation of the Casimir force may then be
done analytically and the sum over all Matsubara frequencies may be
given in a closed form. One
obtains in this manner the following approximation of the Casimir
free energy
\begin{eqnarray}
\label{Eperfectanalytical}
&&\cF^\perf =
-\frac{3 \hbar c R^3}{4 \lambda_\T L^3} \phi(\nu) \quad,\quad
\nu \equiv \frac{2\pi L}{\lambda_\T} \quad, \\
&&\phi(\nu) \equiv \frac{ \nu\sinh\nu + \cosh\nu ( \nu^2 + \sinh^2\nu )}
{2 \sinh^3\nu} ,\quad L \gg R . \nonumber
\end{eqnarray}
The fact that the upper expression is a relevant approximation is
shown on Fig.1: the lower dotted curve, representing the value of
the ratio $\vartheta^\perf$ deduced as in (\ref{deftheta}) through a
derivation of expression (\ref{Eperfectanalytical}), is close to curves computed for small radii $R \ll L$.
Using (\ref{Eperfectanalytical}), it is straightforward to derive an analytical expression
of the entropy $S\equiv-{\partial\cF}/{\partial T}$:
\begin{eqnarray}
\label{Sperfectanalytical}
&&S^\perf = \frac{3 k_\mathrm{B} R^3}{4 L^3}
\left( \phi(\nu) + \nu \phi^\prime(\nu) \right),\quad L \gg R \quad,
\end{eqnarray}
which gives negative values for $\nu\lesssim 1.5$, that is $L\lesssim 1.8\mu$m at $T=300$K.

In addition, we can derive from (\ref{Eperfectanalytical}) low- and high-temperature expressions
for the free energy:
\begin{eqnarray}
\label{perfectlimits}
\cF^\perf & \simeq & 
- \frac{9 \hbar c R^3}{16 \pi L^4} \left( 1 - \frac{\nu^4}{135} + \frac{4\nu^6}{945} \right) 
~ , ~\lambda_T \gg  L \gg R,  \nonumber\\
\cF^\perf & \simeq & -\frac{3 \hbar c R^3}{8 \lambda_\T L^3}
\quad,\quad L \gg \lambda_\T ,\, R .
\end{eqnarray}
Equation (\ref{perfectlimits}) is the large-distance high-temperature limit 
which can be generalized to metallic scatterers described by either the plasma or the Drude model. 
Starting with the lossless plasma model ($\gamma=0$) we obtain for $L\gg\lambda_\P$  Fresnel
coefficients with unit modulus $r_\TE \approx -1,\, r_\TM \approx 1$, 
while the low-frequency expansion of the Mie coefficients \cite{TannerPRB84}, 
and the resulting free energy, are read, introducing the parameter $\alpha\equiv \frac{2\pi R}{\lambda_\P}$:
\begin{eqnarray}
\label{Mieplasma}
&&a_1^\plas \simeq - \frac{2 \txi^3}{3} \;,\;
b_1^\plas \simeq \left( \frac{1}{3}+ \frac{1}{\alpha^2}-
\frac{\coth\alpha}{\alpha} \right) \txi^3 , \\
&&\cF^\plas \simeq  - \frac{3 \hbar c R^3}{8 \lambda_\T L^3}
\left( 1 + \frac{1}{\alpha^2}-\frac{\coth\alpha}{\alpha} \right) ~ ,~   L \gg \lambda_\T, R .
\nonumber
\end{eqnarray}
The result for perfect reflection is reproduced by (\ref{Mieplasma}) when both $L, R \gg \lambda_\P$.

For the dissipative Drude model ($\gamma \neq 0$), the low frequency limit of the two Fresnel coefficients have the
well-known form $r_\TE \to0,\, r_\TM \approx  1$. The low-frequency expansion of the Mie
coefficients and the ensuing free energy are read
\begin{eqnarray}
\label{MieDrude}
&&a_1^\Drud \simeq - \frac{2 \txi^3}{3} + \frac{c \txi^4 }{\sigma_0 R} \;,\;
b_1^\Drud \simeq  \frac{\sigma_0 R \txi^4 }{45c} ,  \nonumber \\
&&\cF^\Drud \simeq  - \frac{\hbar c R^3}{4 \lambda_\T L^3} \quad , \quad  L \gg \lambda_\T, R .
\end{eqnarray}
The long distance free energy for the Drude model amounts to 2/3 of the
value for perfect mirrors whereas this ratio is 1/2 in the
plane-plane geometry. The latter result is explained by the fact
that the TE reflection coefficient vanishes at zero frequency so
that only the TM modes contribute \cite{Bostrom00,Brevik06}. The
change of the ratio 1/2 to 2/3 in the plane-sphere geometry has to
be attributed to the redistribution of the TE and TM contributions
into electric and magnetic spherical eigenmodes.

Formally the results for the Drude model (\ref{MieDrude}) can be
obtained from the plasma model results (\ref{Mieplasma}) by taking
the limit $R\ll\lambda_\P$. In this limit however, we should take
into account the effect of quantum confinement in the small sphere,
which is out of the scope of the present letter. Two further
features in (\ref{MieDrude}) must be emphasized. First, the
coefficient $b_1$ is vanishingly small in the Drude model but not in
the plasma model; the latter can thus not be obtained by turning the
relaxation frequency $\gamma$ to zero (or $\sigma_0$ to $\infty$).
In addition, the free energy for the Drude model is independent of the values of $\lambda_P$ and $\gamma$,
whereas the one for the plasma model depends on $\lambda_P$.

On Fig.3, we illustrate the comparison of the two models by plotting the ratio of
the thermal Casimir forces $F^\plas$ calculated with the plasma model
and $F^\Drud $ obtained with the Drude model.
Again, the plots correspond to $\lambda_\P=136$nm and $\lambda_\gamma/\lambda_\P=250$.
\begin{figure}[t]
\centering
\includegraphics[width=8cm]{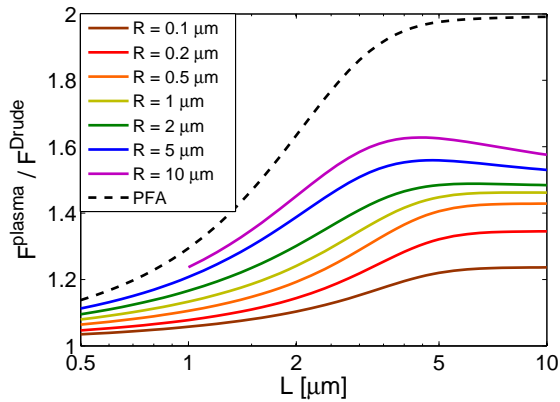}
\caption{Ratio of thermal Casimir force at $T=300K$
calculated with the plasma model and the Drude model,
as a function of surface separation $L$ for different radii of the sphere.
The solid curves from bottom to top correspond
to increasing values of sphere radii.
The dashed curve is the PFA prediction. [Colors online].}
\label{Fig3}
\end{figure}
The results of our calculations are shown by the solid curves with
the sphere radius increasing from bottom to top as in Fig.1. The ratio
$F^\plas/F^\Drud$ varies in the plane-sphere geometry as a function
of the sphere radius, which clearly demonstrates the strong
interplay between the effects of temperature, dissipation and
geometry. For large spheres ($R \gg \lambda_\P$), the ratio converges
at long distances to the value 3/2 which has been obtained analytically in the
preceding paragraphs, whereas it remains smaller for small spheres
(down to $~1.2$ for $R\sim100$nm). The dashed curve gives the
variation of the same ratio as calculated within the PFA which leads
to a factor of 2 in the limits of large distances or high temperatures.
We emphasize that the factor of 2 deduced within PFA is never
approached at the large distance limit 
within the calculations performed in the plane-sphere geometry.

To summarize we have computed exact results for the Casimir free energy and force 
at non zero temperature in the plane-sphere geometry. 
We have given plain evidence for a strong correlation between the effects of
geometry, temperature and dissipation based on the perfect reflector, plasma and 
Drude model to describe material properties. 
The correlation becomes clearly visible in the relative approaching of the Casimir force values 
computed with the Drude and plasma model, the appearance of negative entropies evidently 
not related to the presence of dissipation and the fact that PFA underestimates the Casimir force 
for the Drude model at short distances while it overestimates it for the plasma model. 
If the latter feature were conserved for the experimental parameter region $R/L$ $(>10^2)$, 
the actual values of the Casimir force calculated within plasma and Drude model could turn 
out to be closer than what PFA suggests, eventually diminishing the discrepancy between 
experimental results and predictions of the thermal Casimir force using the Drude model. 
Settling this question is an open and highly topical program in Casimir physics.

\acknowledgements The authors thank G.-L. Ingold for many fruitful
discussions, CAPES-COFECUB and the French Contract
ANR-06-Nano-062 for financial support, and the ESF Research Networking Programme CASIMIR
(www.casimir-network.com) for providing excellent opportunities for
discussions on the Casimir effect and related topics. P.A.M.N. thanks CNPq and Faperj for financial support.

\newcommand{\REVIEW}[4]{\textrm{#1} \textbf{#2}, #3, (#4)}
\newcommand{\Review}[1]{\textrm{#1}}
\newcommand{\Volume}[1]{\textbf{#1}}
\newcommand{\Book}[1]{\textit{#1}}
\newcommand{\Eprint}[1]{\textsf{#1}}
\def\etal{\textit{et al}}

\end{document}